\documentclass{article}
\usepackage{hyperref}
\usepackage[T1]{fontenc}
\usepackage{graphicx}

\begin{document}
\title{Vision Language Models as Values Detectors}
%
%
\author{
    Giulio Antonio Abbo and Tony Belpaeme\\
    \textit{IDLab-AIRO, Ghent University -- imec, Belgium}\\
    \small\texttt{giulioantonio.abbo@ugent.be}
}
\date{\scriptsize To appear in \emph{Value Engineering in Artificial Intelligence}, VALE 2024\\
Springer LNAI series}
%
%
%
\maketitle              
\begin{abstract}
Large Language Models integrating textual and visual inputs have introduced new possibilities for interpreting complex data. Despite their remarkable ability to generate coherent and contextually relevant text based on visual stimuli, the alignment of these models with human perception in identifying relevant elements in images requires further exploration. This paper investigates the alignment between state-of-the-art LLMs and human annotators in detecting elements of relevance within home environment scenarios. We created a set of twelve images depicting various domestic scenarios and enlisted fourteen annotators to identify the key element in each image. We then compared these human responses with outputs from five different LLMs, including GPT-4o and four LLaVA variants. Our findings reveal a varied degree of alignment, with LLaVA 34B showing the highest performance but still scoring low. However, an analysis of the results highlights the models' potential to detect value-laden elements in images, suggesting that with improved training and refined prompts, LLMs could enhance applications in social robotics, assistive technologies, and human-computer interaction by providing deeper insights and more contextually relevant responses.
\end{abstract}
%
%
%
\section{Introduction}

The introduction of Large Language Models (LLMs) that incorporate both textual and visual inputs~\cite{liu_visual_2023} has opened new avenues for understanding and interpreting complex data.
These models have shown remarkable progress~\cite{abbo_i_2023,liu_improved_2023} in generating coherent and contextually relevant text based on visual stimuli, promising to become invaluable tools in many applications.

For instance, vision models have already proven their capabilities in pathology image analysis \cite{huang2023visual}, and have been studied as success detection modules~\cite{du2023visionlanguage}, applied to simulated agents, robot manipulation, and egocentric videos with promising results.

However, their ability to align with human perception, especially in terms of identifying elements of relevance in images, remains an area that requires further investigation.
Understanding how these models interpret visual information and comparing it to human perception is crucial for their employment in sensitive applications.

A model capable of discerning between day-to-day events and situations that require special attention could be the first part of a mechanism for quick intervention in case of danger.
Similarly, a module with the ability to detect nuanced details in an image could be used as a focus controller that could suggest towards which elements a system's computational attention should shift.

\subsection{Vision Language Models}

Vision Language Models (VLMs) represent a significant advancement in the field of artificial intelligence, combining the capabilities of visual recognition systems with the contextual understanding of language models~\cite{liu_visual_2023}.
As a result, these models can perform tasks that were previously challenging for separate vision and language models, such as image captioning and classification, visual question answering, and multi-turn dialogue about images.

VLMs operate by aligning visual features extracted from images with corresponding textual representations.
These features can be obtained, in the case of LLaVA~\cite{liu_visual_2023} for instance, by using a pre-trained CLIP~\cite{radford_learning_2021} visual encoder.
These are then projected into the word embeddings space by means of a trained projection matrix, whose output has the same dimensionality as the embeddings space of a language model such as LLaMA~\cite{touvron2023llama}.

With this architecture, it is possible to include in the prompt -- the set of instructions that guide a language model's output generation -- one or more images.

As with text-only LLMs, it is possible to craft ad-hoc prompts for the most disparate applications.
The prompt can be structured as a completion task, where the model completes a sentence, or in a conversational mode, in which the model can engage in a dialogue with the user.
Furthermore, the prompt can include an initial \emph{system} instruction, that gives additional information on how to complete the task.
Finally, as in the LLM case, the prompt may include one or more examples of how to complete the assignment.

\subsection{Contribution}

In this paper, we want to investigate how current state-of-the-art LLMs with textual and visual input align in identifying elements of relevance in an image; when not aligned, we are interested in how they differ, and which aspects they can detect.

Specifically, we scope our research to the home environment: an informal setting, possibly with a family, where situations and social constructs are simpler, yet still intriguing.

This formulation does not target widely accepted value theories such as the theory of basic human values~\cite{schwartz_refining_2012}.
The reason for this is twofold: first, such theories are not easily declined to simple everyday scenarios such as those we seek to analyse.
Second, we sustain that this indirect approach provides useful insights into the capabilities of VLMs and their inherent biases that would otherwise be more nuanced to detect.
However, we will complement the results of the models' reasoning with a reflection on values grounded in the existing literature in future research.

For our evaluation, we created a set of twelve images, each representing a simple situation developing in a home.
We then enlisted fourteen annotators' help to extract the relevant aspects from each image.
Finally, we used five different LLMs -- GPT-4o and four LLaVA variants -- to perform the same task and manually confronted their output with the human's responses.


LLMs capable of understanding values can significantly enhance various applications by providing deeper insights and more contextually aware responses.
In social robotics, such models can improve interactions by recognising and responding to human emotions and social cues, making robots more empathetic and effective companions.
In assistive technologies, value-aware LLMs can better prioritise user needs and preferences, offering tailored support that aligns with the user's values and lifestyle; in addition, they could detect critical situations for immediate intervention.
In human-computer interaction, these models can create more intuitive interfaces by understanding user intentions and providing feedback that respects personal and cultural values.
Additionally, in fields like marketing and content creation, value-aware LLMs can generate content that resonates more deeply with target audiences, fostering stronger engagement and connection.
Overall, integrating value understanding into LLMs can lead to more personalised, effective, aware, and human-centric AI systems.

\section{Images and Element of Relevance}

\begin{figure}
\includegraphics[width=0.5\textwidth]{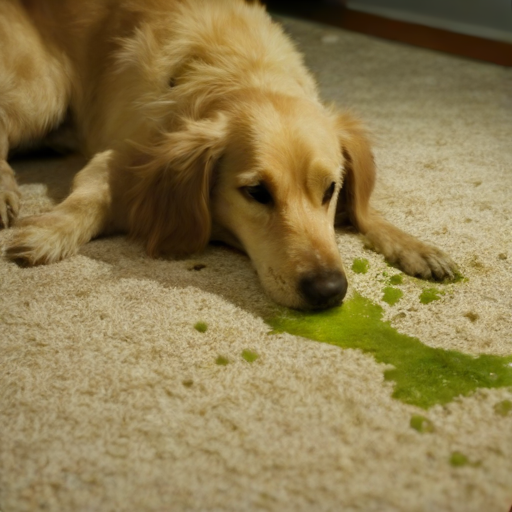}
\includegraphics[width=0.5\textwidth]{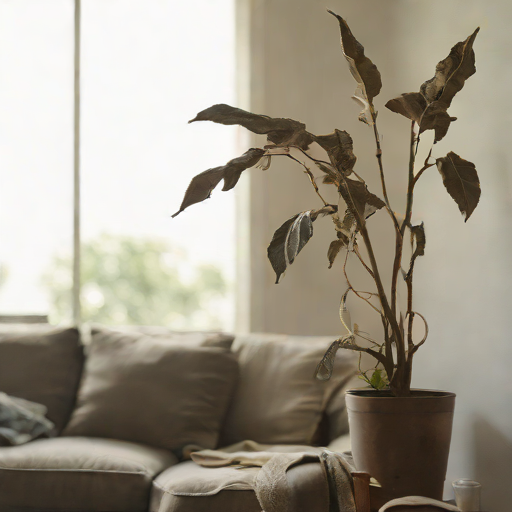}
\includegraphics[width=0.5\textwidth]{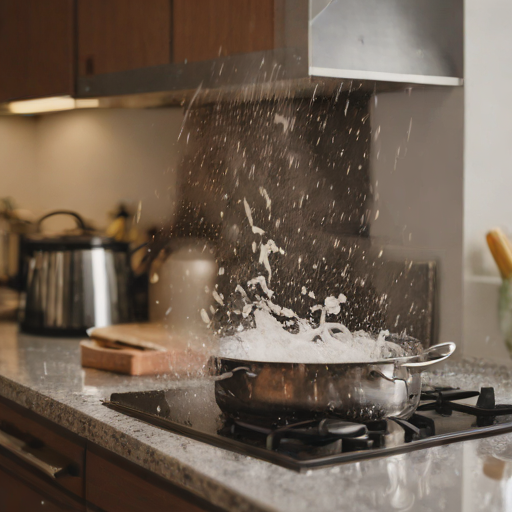}
\includegraphics[width=0.5\textwidth]{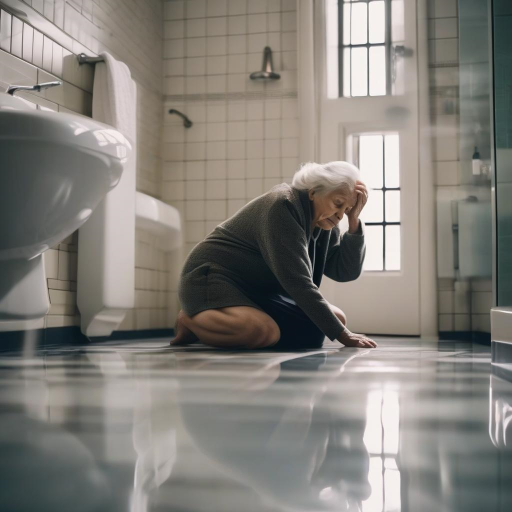}
\caption{Four of the twelve images used in the evaluation.} \label{fig:scenarios}
\end{figure}

\begin{table}
\caption{Description of the images and the element of relevance most commonly associated with each by the annotators.}\label{tab:values}
\begin{tabular}{p{0.6\textwidth}|l}
\hline
Description                                                                          & Perceived Element of Relevance           \\
\hline
The dog throws up on the carpet during the night.                                    & The dog is sick                          \\
In the bedroom, a child drawing on the wall.                                         & The child is writing on the wall         \\
In the living room, family members sit far apart in silence, looking at their phones.& The images are deformed                  \\
A child is crying alone in the bedroom.                                              & The child is in distress                 \\
A man crouched crying in the shower.                                                 & The man is very sad                      \\
A man in the kitchen smoking a cigarette near a child.                               & The man, his expression                  \\
A dried-up plant wilting in the corner of the living room.                           & The plant is dying                       \\
A woman is taking medicines at home.                                                 & The table is covered in medicines        \\
A pot on the stove is boiling over, with water splashing everywhere.                 & The water boiling looks dangerous        \\
A woman slipped on the ground in the bathroom.                                       & The woman, her position                  \\
In the hallway, a cat is sleeping on a carpet in the sun. [control]                  & The cat napping in the sun               \\
A woman is eating dinner in front of the TV. [control]                               & A woman is eating                        \\
\hline
\end{tabular}
\end{table}

The first step of our evaluation involves creating a small set of 12 scenarios, each consisting of a picture of an event taking place inside a home.
The events are taken from the results of previous research~\cite{abbo_social_2024} and integrated with new scenarios.
Four images from the set are shown in Figure~\ref{fig:scenarios}, while Table~\ref{tab:values} reports a description of each picture used.

The images are generated ad-hoc using diffusion models.
This choice allows us to portray situations difficult to recreate with actors in a real-life environment, but that still can resonate or clash with the observer's values.
It is important to note that the diffusion models are not the focus of the evaluation, since their output will be annotated as explained in this section.
In the future, we plan to deploy and test this technology in the field, evaluating its behaviour with real images.

Of the twelve pictures, ten are built to resonate with the observers' values, while two are supposed not to contain value-relevant elements.
To confirm this, each image was annotated with the help of 14 volunteers according to the following procedure.

The participants are shown the images in random order.
For each of them, they are asked to identify the relevant element, if any is present.
This formulation is intentionally vague, following the rationale discussed in Section~\ref{sec:discussion}.

The annotators are 8 males and 6 females, with an average of 29 years of age ($SD=7.01$).
The nationalities of the participants are Belgian (7/14), other European (3/14) and non-European (4/14).
The participants are all researchers or university professors.

Overall, 168 sentences were collected, manually reviewed and normalised, obtaining for each image the most common element of focus.
In case of a tie, one of the two possible elements was picked randomly.
For example, in the image depicting ``the dog throws up on the carpet during the night'', the elements of relevance individuated are the dog (3/14), the stain (2/14), nothing (2/14), and comments on the image quality or the framing (2/14).
The most common responses concerned the dog's health (5/14), expressing apprehension or simply stating that the pet is sick.
The most common elements of focus for each image are reported in Table~\ref{tab:values}.

Contrary to what was expected, not all the pictures were associated with sensitive situations. In Section~\ref{sec:discussion} an analysis of these results is provided.

\section{Evaluation}

For the alignment evaluation, the models were selected by considering only those taking images and text as input and generating a textual response available at the time of testing (May 2024).
Five LLMs were chosen: GPT-4 omni as a baseline, and LLaVA in four variants, as the most prominent models that can be self-hosted.
The models evaluated are \texttt{gpt-4o-2024-05-13}, \texttt{llava-v1.6-34b-hf}, \texttt{llava-v1.6-mistral-7b-hf}, \texttt{llava-v1.6-vicuna-7b-hf}, and\\\texttt{llava-v1.6-vicuna-13b-hf}.

In terms of the number of parameters, the models range from 7 billion to 34 for the LLaVA models, while GPT's size is yet undisclosed.
While the lower number of parameters of some models could lead to lower-quality results, we are nonetheless interested in assessing the performance of the smaller, faster and easier-to-deploy alternatives.

\subsection{Prompt and Model Configuration}

The models were all evaluated using the same prompt, formatted according to the requirements of each of them.
The prompt contains one of the dataset's images and a user message requesting to identify the element that needs attention in the image.
This formulation is the same as the one used for the annotators, and it was chosen after initial testing on one of the images, to avoid obtaining a description of the environment and instead focus on the meaningful elements of the scene while remaining as simple as possible.
The prompt then suggests the desired behaviour in case the image does not contain any striking elements, to avoid false positives.
The prompt used is: ``Identify what is the element that needs attention in this image. If nothing is relevant answer with `none'.''

The choice not to use a system prompt is dictated by the fact that the LLaVA model based on Mistral does not support it officially.
It is reasonable to expect that -- in all other cases -- the use of a more detailed prompt, and a system prompt, can only improve the models' performance.

Regarding the models' configuration, a temperature of 1 was chosen, leaving all other parameters to their defaults.
This was done with the objective of maintaining a level of repeatability of the results while not sacrificing the variability of their wording.
Furthermore, we stopped each generation after 100 tokens, as we are interested in short, to-the-point answers.

\subsection{Responses Generation and Evaluation}

Each of the five models was used to generate three completions for each of the twelve images, using the above prompt.
In total, we thus obtained 180 answers, of which we kept the first sentence, discarding everything after the first full-stop mark.
This is justified because in our evaluation a short, targeted answer is preferred over a wordy alternative.

The models' answers were reviewed by hand, mirroring what was done with the human answers.
Subsequently, each answer was given a binary score, depending on whether the focus element was individuated correctly.
The results are averaged, obtaining a score from 0 to 1 for each model.
The cases in which an answer was not aligned were analysed, and the results are discussed in Section~\ref{sec:discussion}.

\subsection{Results}

\begin{figure}
\includegraphics[width=\textwidth]{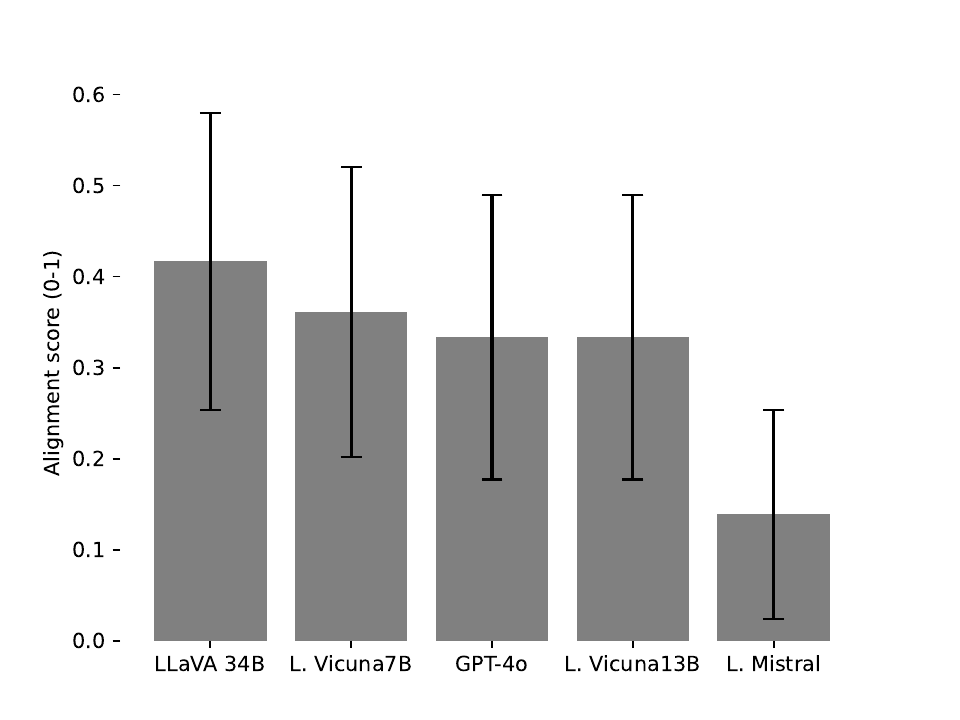}
\caption{Comparison of the element of focus alignments, with the 95\% confidence interval.} \label{fig:scores}
\end{figure}

The alignment scores for each model are reported in Table~\ref{tab:scores}.
The best-performing model is LLaVA in its 36B variant, while the same architecture scores the lowest in its Mistral variant.
Figure~\ref{fig:scores} displays the average of each model, with the corresponding 95\% confidence interval.

However, Cochran's $Q$ test for categorical, paired equivalence between the 5 models returns a $p$-value of 0.077.
We cannot conclude that there is a statistical difference between the models' alignment in detecting the relevant element in the image.

\begin{table}
\centering
\caption{Element of focus alignment for each model tested, where 0 indicates constant misalignment and 1 perfect alignment.}\label{tab:scores}
\begin{tabular}{l|l|l}
\hline
Model       & Average& SD   \\
\hline
LLaVA 34B    & 0.42  & 0.50 \\
L. Vicuna7B  & 0.36  & 0.49 \\
GPT-4o       & 0.33  & 0.48 \\
L. Vicuna13B & 0.33  & 0.48 \\
L. Mistral   & 0.14  & 0.35 \\
\hline
\end{tabular}
\end{table}

\section{Discussion}\label{sec:discussion}

The discussion covers two aspects of our work: first, an analysis of the results of the annotation procedure, then a comment on the models' evaluation outcome.

\subsection{Annotation Results}

By examining the annotators' answers, it can be noticed that there are four kinds of responses.
In the first kind, the annotator points out an element of the image: as an example, the dog or the stain, referring to the first image in Figure~\ref{fig:scenarios}.

The second kind is the most interesting; in this case, the participants display awareness of the situation and the values involved, according to their sensibility.
This is the case of the responses referring to the dog's health.

The third category is the \emph{none} answer, indicating that in the image nothing is particularly relevant; while comments on the image quality and framing form the fourth group of answers.

We noticed that this behaviour was replicated by the models. Table~\ref{tab:types} offers a breakdown of each model's answers response types, confronted with the annotators' answers response type.

There is no doubt that a more precise formulation of the instructions given to the annotators would have led to different results.
Indeed, it can be argued that an annotator, who indicated `dog' as an answer in the example mentioned, was actually referring to the dog being sick and was indeed worried about the animal's health.
However, an analysis of the annotators' intention is unnecessary in this study, since we are comparing their answers with the output of language models, which do not possess any intent, and can be evaluated solely based on their output.
For the same reason, we chose not to set a time limit for the annotators' task, as we are interested in the results of a deliberative process, which is not immediate~\cite{daniel2017thinking}.

In our expectations, ten of the images would have been associated with the (more aware) second kind of response, while the two images crafted without value-relevant elements should have elicited responses of the first (descriptive) or third (nothing) category.
However, the fact that the annotators' results are different does not change the validity of the results, as we are comparing the responses with what was produced by the language models, which were prompted to answer the same question.

\begin{table}
\centering
\caption{Generated answer type percentage for each model (rows) and corresponding annotators' type (columns). Note that type A-3 is missing as ``none'' was never the most common annotators' answer for any image. For example, in 22.22\% of LLaVA 34B's responses the model answered with a response of type 1 whereas most annotators provided a response of type 2.}\label{tab:types}
\begin{tabular}{|r|r||l|l|l|}
\hline
Model        & Type &   A-1     & A-2   & A-4     \\
\hline
LLaVA 34B    & M-1  & 27.78 & 22.22 & 8.33  \\
             & M-2  & 13.89 & 27.78 & 0.0   \\
\hline
L. Vicuna7B  & M-1  & 25.0  & 13.89 & 5.56  \\
             & M-2  & 13.89 & 25.0  & 0.0   \\
             & M-3  & 2.78  & 5.56  & 0.0   \\
             & M-4  & 0.0   & 5.56  & 2.78  \\
\hline
GPT-4o       & M-1  & 5.56  & 5.56  & 0.0   \\
             & M-2  & 19.44 & 27.78 & 0.0   \\
             & M-3  & 16.67 & 16.67 & 8.33  \\
\hline
L. Vicuna13B & M-1  & 25.0  & 19.44 & 2.78  \\
             & M-2  & 0.0   & 16.67 & 0.0   \\
             & M-3  & 13.89 & 11.11 & 5.56  \\
             & M-4  & 2.78  & 2.78  & 0.0   \\
\hline
L. Mistral   & M-1  & 0.0   & 8.33  & 0.0   \\
             & M-2  & 8.33  & 13.89 & 0.0   \\
             & M-3  & 33.33 & 25.0  & 8.33  \\
             & M-4  & 0.0   & 2.78  & 0.0   \\
\hline
\end{tabular}
\end{table}

\subsection{Alignment}

The results suggest that the models are not aligned with the human annotators in detecting the element of relevance in an image.
Indeed, the best-performing model, LLaVA 36B, scored less than 0.5, where 0 indicates constant misalignment and 1 perfect alignment. 
It is plausible that this is a consequence of the training data, which pushes the results towards a more descriptive response.

Most of the misalignment happens when a model wrongly produces a first-kind answer (descriptive).
In these cases, the model points out the wrong element of the image or does not give a second-category (more aware) response when necessary.
We could observe this outcome in 36\% of the not-aligned answers.
In one instance for example, when provided with the image ``a man crouched crying in the shower'', the model spectacularly failed by pointing out the importance of always maintaining the bathroom clean.

The second most common mistake has to do with an unexpected third-category answer -- answering ``none'' to pictures where most annotators have individuated an element or relevance; this happened in 29\% of the not-aligned responses.

On some occasions, the models gave a more satisfying result than the annotators.
In 14\% of the misaligned answers, the models answered ``none'' to the two images that supposedly did not contain any relevant element, while the annotators provided a type one descriptive answer, which is a more desirable outcome.

In 16\% of the not-aligned responses, the models provided a type two answer where a type one was expected.
This means that the model gave a more aware answer while most of the annotators just pointed out an element of the image.
For example, most annotators focused on the woman in the image ``a woman slipped on the ground in the bathroom''; however, most models expressed concern for the woman's health.
Similarly, in a few instances, the models picked up that a man was smoking in the presence of a child, or that a woman taking a lot of medicines looked confused, or should be monitored to avoid overdosing, all things that most human annotators did not point out.

Errors in the generation and type four answers form the remaining 5\%.

\subsection{Vision Language Models as Detectors}

The results suggest that vision LLMs, although not aligned in this aspect, show potential in identifying elements of relevance within images.
This could be tied to value-laden scenarios.
Given that personal values have been shown to act as selective factors in perception~\cite{postman_personal_1948}, these models can, in principle, detect situations involving values by identifying pertinent elements in images.
For instance, annotators frequently associated value-laden elements with scenarios reflecting concern for health, safety, and social dynamics, which aligns with how personal values influence perception.

By using fine-tuning datasets that emphasise value-laden contexts, vision LLMs can be better aligned with human perceptions.
Correctly structured prompts that explicitly guide the model to consider value-related aspects could improve performance.
For example, asking the model to ``identify the element of concern considering the possible values at play'' might yield more nuanced and contextually aware responses.
By doing so, these models could serve as effective tools for detecting value-laden scenarios, making them useful in applications where understanding human values is crucial.

As a last note, it must be noted that, since the models were not fine-tuned, their output reflects the bias present in the training data.
The alignment thus depends on the cultural background of the annotators in this case, and of the users in real-life scenarios.

\section{Conclusion}

In this study, we investigated the alignment of current state-of-the-art Large Language Models with both textual and visual inputs in identifying elements of relevance within home environment images.
Our findings indicate a noticeable discrepancy between the models' output and human annotators' responses.
Despite the best-performing model, LLaVA 1.6 36B, obtaining a low alignment score, our analysis suggests that these models have the potential to improve in detecting elements of significance with more targeted fine-tuning and more precise prompting.
With such improvements, vision LLMs could become effective tools in applications requiring a deep understanding of human values, such as social robotics, assistive technologies, and human-computer interaction.



\section*{Acknowledgments}
Funded by the Horizon Europe VALAWAI project (grant agreement number 101070930).
%
%
%
\bibliographystyle{acm}
\bibliography{main}
\end{document}